\documentclass[aps, twocolumn, nofootinbib, superscriptaddress, prd]{revtex4-2}
\usepackage{pkgs}
\usepackage{booktabs}

\begin{document}
\title{Time-Domain Axion Searches with Magnetic White Dwarfs}

\author{Hong-Yi Zhang}
\thanks{These authors contributed equally to this work.}
\affiliation{School of Physics, Xi'an Jiaotong University, Xi'an 710049, China}
\affiliation{Tsung-Dao Lee Institute \& School of Physics and Astronomy, Shanghai Jiao Tong University, Shanghai 201210, China}
\author{Heng Bian}
\thanks{These authors contributed equally to this work.}
\author{Zhao-Yu Zuo}
\thanks{Contact author: zuozyu@xjtu.edu.cn}
\affiliation{School of Physics, Xi'an Jiaotong University, Xi'an 710049, China}

\date{\today}

\begin{abstract}
Magnetic white dwarfs can convert photons into axions in their strong magnetic fields, with the conversion probability modulating the light curve as the star rotates. However, this observable is degenerate with intrinsic stellar variability if the background is modeled too simplistically. We develop a controlled background-degeneracy framework that computes the axion-induced modulation from reconstructed stellar magnetic fields while fitting it simultaneously with a flexible Fourier model of the intrinsic light curve. Applying this framework to TESS observations of PG~1015+014 using two independent magnetic field reconstructions, we find that a sinusoidal stellar background can produce an apparent preference for nonzero axion-photon conversion. This preference is absorbed once the background light curve includes the second harmonic, indicating that higher-harmonic stellar variability is a leading degeneracy for precise photometric axion searches in magnetic white dwarfs. Interpreting the two-harmonic fit as a conservative baseline, we obtain competitive constraints for sub-$\mu\mathrm{eV}$ axions. We further derive an analytic target-ranking estimate for other TESS magnetic white dwarfs, identifying systems where phase-resolved magnetic modeling would be most valuable for competitive axion probes.
\end{abstract}
\maketitle

\section{Introduction}
Dark matter accounts for about $27\%$ of the energy density of the universe \cite{Planck:2018vyg, ParticleDataGroup:2024cfk}, but its microscopic identity remains unknown \cite{Cirelli:2024ssz}. Axions are among the best motivated candidates: the QCD axion arises from the Peccei--Quinn solution to the strong CP problem \cite{Peccei:1977hh, Weinberg:1977ma, Wilczek:1977pj}, while axionlike particles appear more broadly in extensions of the Standard Model and need not obey the QCD axion mass-coupling relation \cite{Marsh:2015xka, DiLuzio:2020wdo}. The viability and phenomenology of axions as dark matter candidates have been extensively studied in cosmology and astrophysics \cite{Edwards:2020afl, Zhang:2023ktk, Salehian:2021khb, Amin:2020vja, Zhang:2023vva, Walters:2024vaw, Yin:2025amn, Zhang:2025pgb, Cheek:2025kks, Zhang:2024bjo, Amin:2019ums, Yin:2024xov, Bhura:2026bpy}. A central experimental target is the axion-photon interaction,
\begin{align}
\label{lagrangian}
\mcal{L}_{a\gamma}=g_{a\gamma}a\,\bE\cdot\bB ~,
\end{align}
which mixes photons with axions in an external magnetic field \cite{Raffelt:1987im}. Because the conversion probability grows with magnetic-field strength and coherent propagation length, compact magnetized objects provide sensitive environments for probing small values of $g_{a\gamma}$.

Magnetic white dwarfs (MWDs) are especially useful in this regard. Their surface fields can reach $10^6$--$10^9\,\mrm{G}$, their radii provide macroscopic baselines, and many nearby systems are bright enough for precision optical photometry, spectroscopy, and polarimetry \cite{Ferrario:2015oda}. Earlier work proposed using radiation from magnetic stars and MWDs to constrain axion-photon conversion \cite{Lai:2006af, Gill:2011yp}. This program has recently developed into several observational channels: X-ray searches for axions produced in MWD interiors and converted in magnetospheres \cite{Dessert:2019sgw, Dessert:2021bkv, Ning:2024ozs}, and optical spectropolarimetric searches for axion-induced linear polarization in thermal surface emission \cite{Dessert:2022yqq,Benabou:2025jcv}. These measurements probe related conversion physics but differ in their observables and systematics, making MWDs a useful multichannel setting for sub-$\mu\eV$ axion searches.

\begin{figure}
\centering
\includegraphics[width=\linewidth]{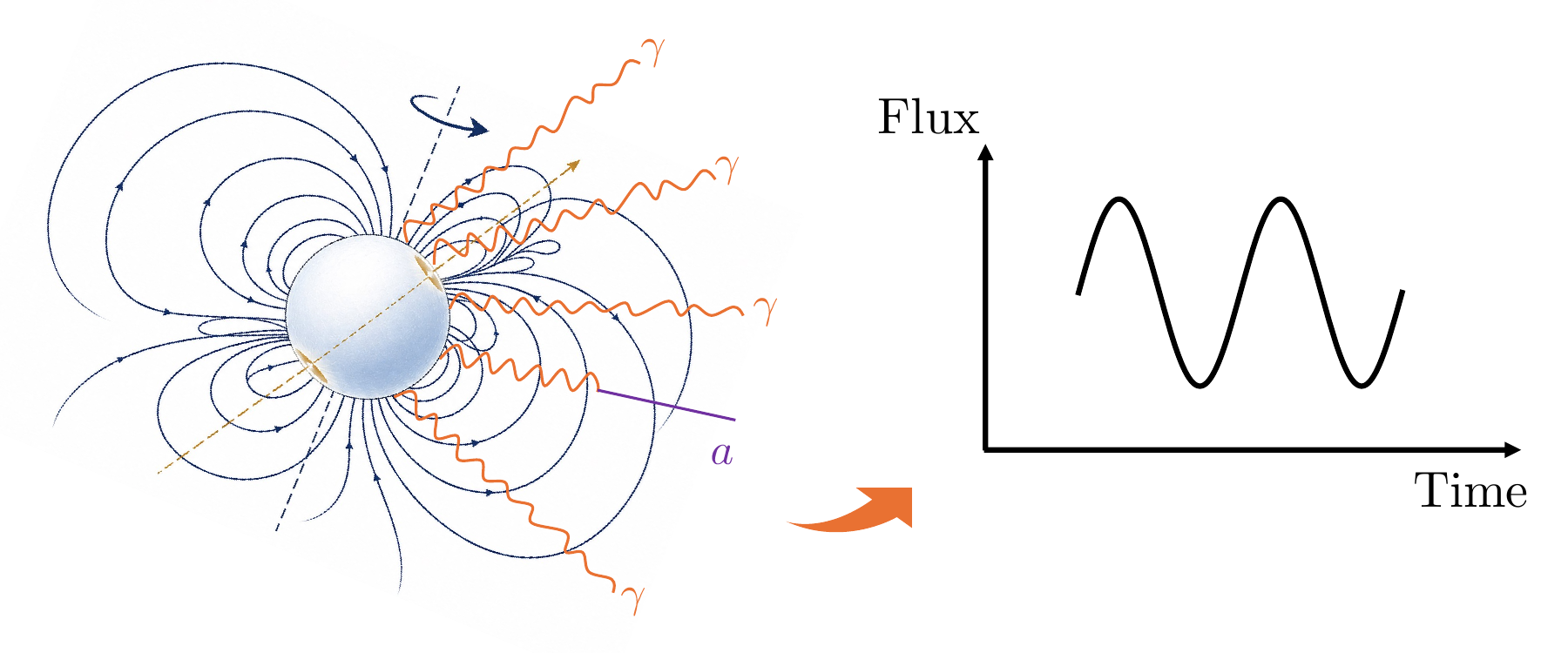}
\caption{Magnetic white dwarfs emit photons from the surface, a small fraction of which can convert to axions in the surrounding strong magnetic fields. The resulting phase-dependent attenuation can be searched for in light curve data, subject to degeneracy with intrinsic stellar variability.}
\label{fig:mwdcartoon}
\end{figure}

A complementary observable is time-dependent flux modulation. As a MWD rotates, the transverse magnetic field encountered by photons emitted from the visible stellar surface changes with rotational phase. The photon survival probability therefore becomes phase dependent, imprinting a periodic modulation on the light curve. See figure \ref{fig:mwdcartoon} for an illustration. This possibility is timely because missions such as the Transiting Exoplanet Survey Satellite (TESS) \cite{2015JATIS...1a4003R} provide high-cadence optical light curves for many rotating MWDs, while Zeeman tomography can reconstruct nondipolar magnetic topologies for benchmark systems \cite{Euchner:2002qv, Euchner:2006ku}. A recent light curve study applied this idea to PG 1015+014 and showed that photometric variability can be sensitive to axion-photon conversion by assuming a sinusoidal background light curve \cite{Tian:2025lkp}.

However, MWD photometry is not a clean axion observable by itself. Rotational light curves can contain intrinsic stellar structure beyond a purely sinusoidal background from magnetic dichroism, surface-brightness inhomogeneities, phase-dependent Zeeman opacity, limb-darkening effects, and nondipolar magnetic fields \cite{2013ApJ...773...47B, Maxted:2000wp, Valyavin:2011rb, Maoz:2014upa, 2024arXiv240115158H}. If the stellar background is too restrictive, residual phase structure can be incorrectly assigned to axion-photon conversion. As we will see, this could indeed be the case for precision light curve data if a sinusoidal background is assumed. Therefore, a time-domain axion search requires a framework that treats the axion-photon prediction and the intrinsic stellar variability on the same statistical footing.

In this work we develop a phase-resolved photometric framework designed to make this degeneracy explicit, allowing for a systematic approach to probe axions with precision light curve data. For a given magnetic-field reconstruction, we compute the surface-averaged photon survival probability as a function of rotational phase and fit the resulting axion-induced modulation together with a flexible phenomenological Fourier model for the intrinsic stellar light curve. The Fourier coefficients are nuisance parameters and are marginalized over in the statistical analysis. The relevant observable is therefore not merely the amplitude of the rotational modulation, but the phase-dependent structure that remains distinguishable from ordinary stellar harmonics after background marginalization. This framework is general and can be applied to any rotating magnetized star where the magnetic geometry is known or reconstructable.

We apply the framework to PG 1015+014, a rapidly rotating MWD with period $98.7\,\mrm{min}$ \cite{1988MNRAS.235.1451W, 2013ApJ...773...47B} and phase-resolved magnetic reconstructions from spectroscopic and circular polarization data \cite{Euchner:2006ku}. We analyze TESS observations of this target using both the three off-centered dipole model and the truncated multipole model for the magnetic field. We find that both reconstructions give compatible constraints on $g_{a\gamma}$, indicating that the derived axion constraint is not tied to one particular parametrization of the surface field. We also derive a target-ranking projection for additional nine MWDs. While a more careful object-specific analysis is required, our estimates suggest that MWD photometry could lead to competitive sensitivity to the axion-photon coupling even with existing TESS data.

The rest of this work is organized as follows. We first describe axion-photon conversion in rotating MWD magnetospheres in Sec.~\ref{sec:axion_photon_conversion}. We then introduce the magnetic field reconstructions in Sec.~\ref{sec:magnetic_photometric_modeling}, the TESS data reduction and phase-folding procedure in Sec.~\ref{sec:tess}, and the phenomenological Fourier model for the intrinsic stellar light curve in Sec.~\ref{sec:background_model}. The statistical framework used to constrain the axion parameters and compare competing models is presented in Sec.~\ref{sec:bayesian_constraints}. Finally, we summarize and discuss the results in Sec.~\ref{sec:conclusions}.

\section{Axion-photon conversion in rotating magnetic white dwarfs}
\label{sec:axion_photon_conversion}
The axion-photon interaction \eqref{lagrangian} mixes the axion with the photon polarization parallel to the magnetic-field component transverse to the photon trajectory. For a rotating MWD, this mixing is intrinsically phase-dependent, since photons emitted from different surface elements propagate through different magnetospheric field profiles as the star rotates. The photometric observable is therefore a phase-dependent photon survival probability that multiplies the intrinsic stellar light curve.

\subsection{Axion-photon conversion probability}
For a photon of angular frequency $\omega$ emitted from a surface element at $\b r$ and propagating along the line of sight $\hat{\b n}$, we write the ray as $\b x(s)=\b r+s\hat{\b n}$, where $s$ is the distance from the stellar surface. In the small-mixing limit, and neglecting plasma and nonlinear QED corrections that are subdominant for the optical MWD systems considered here \cite{Dessert:2022yqq}, the axion-conversion probability is \cite{Raffelt:1987im}
\begin{align}
\label{Pagamma}
P_{a\gamma}(\b r,\f;\omega)
=
\frac{1}{4}g_{a\gamma}^2
\abs{
\int_0^\infty
B_\mrm{T}(s;\b r,\f)
\exp\l(i\frac{m_a^2}{2\omega}s\r)
ds
}^2 ~.
\end{align}
Here $B_\mrm{T}=\abs{\bB-(\bB\cdot\hat{\b n})\hat{\b n}}$ is evaluated along the photon path, and its dependence on the rotation phase $\f$ is fixed by the stellar magnetic topology and viewing geometry. The phase factor in \eqref{Pagamma} describes the loss of coherence when the axion oscillation length becomes shorter than the magnetic-field scale. Taking this scale to be $R_\star$, coherent conversion requires
\begin{align}
\label{light_axion_mass}
m_a
\ll
2\times10^{-7}\,\eV
\l(\frac{\omega}{1\,\eV}\r)^{1/2}
\l(\frac{0.01\RSun}{R_\star}\r)^{1/2} ~.
\end{align}
In this light-axion regime, the leading conversion probability is controlled by the line-of-sight magnetic field integral. A useful estimate follows by taking
$B_\mrm{T}(s)\sim B_0[R_\star/(R_\star+s)]^3$, which gives an integral of order $B_0R_\star/2$. Then
\begin{align}
\label{Pagamma_estimate}
P_{a\gamma}
&\sim
\l(\frac{1}{4}g_{a\gamma}B_0R_\star\r)^2
\nonumber\\
&\sim
0.03
\l(\frac{g_{a\gamma}}{10^{-11}\,\GeV^{-1}}\r)^2
\l(\frac{B_0}{100\,\mrm{MG}}\r)^2
\l(\frac{R_\star}{0.01\RSun}\r)^2 ~.
\end{align}
This scaling explains why MWDs are sensitive despite the small coupling: megagauss fields act over white-dwarf radii, while the rotation converts the spatial magnetic structure into a time-domain signal. For larger $m_a$, the phase in \eqref{Pagamma} varies appreciably across the magnetosphere, leading to suppression in the conversion amplitude.

Equation \eqref{Pagamma_estimate} is a ray-level estimate. In practice, the surface-averaged observable signal is reduced because rays from different surface elements experience different transverse magnetic fields. For PG 1015+014, we find that the surface-averaged conversion probability is smaller than the ray-level estimate by approximately one order of magnitude. This reduction is quantified by the factor $x$ introduced in section \ref{other_MWD}.

\subsection{Surface-averaged flux modulation}
We now connect the ray-level conversion probability to the measured flux. Let $\hat{\b n}$ be the line-of-sight direction, $\mu=\hat{\b n}\cdot\hat{\b r}$ the projection factor, and $D$ the distance to the star. Suppressing the frequency label, the observed flux at rotation phase $\f$ is
\begin{align}
\label{Fobs}
F_\mrm{obs}(\f)
=
\frac{1}{D^2}
\int_{\mu>0}
I_\mrm{S}(\mu,\b r;\f)
P_{\gamma\gamma}(\b r,\f)
\,\mu\,dS ~,
\end{align}
where $I_\mrm{S}$ is the surface specific intensity, $P_{\gamma\gamma}$ is the photon survival probability, and $dS=R_\star^2d\Omega$. For initially unpolarized radiation, only one linear polarization mixes with the axion. To leading order in the conversion probability,
\begin{align}
P_{\gamma\gamma}(\b r,\f) \simeq 1-\frac{1}{2}P_{a\gamma}(\b r,\f) ~.
\end{align}
For the optical photometry used below, we describe the angular dependence of the emergent intensity with a linear limb-darkening law,
\begin{align}
I_\mrm{S}(\mu,\b r;\f) = I_\mrm{S}(1,\b r;\f)f_\mrm{LD}(\mu) \, ,~
f_\mrm{LD}(\mu) = 1-u(1-\mu) ~.
\end{align}
This form follows from a linear source function in radiative transfer \cite{2010eapp.book.....S} and provides an effective description for MWD spectral modeling \cite{Euchner:2002qv}. The coefficient $u$ is assumed to be independent of frequency across the observed band.

It is useful to factor the flux into the intrinsic stellar light curve and a surface-averaged survival probability,
\begin{align}
\label{Fobs1}
F_\mrm{obs}(\f) = F_\mrm{bkg}(\f) \la P_{\gamma\gamma}\ra_\mrm{S}(\f) ~,
\end{align}
with
\begin{align}
\la P_{\gamma\gamma}\ra_\mrm{S} = \frac{ \int_{\mu>0} I_\mrm{S}(\mu,\b r;\f) P_{\gamma\gamma}(\b r,\f) \,\mu\,dS }{ \int_{\mu>0} I_\mrm{S}(\mu,\b r;\f) \,\mu\,dS } ~.
\end{align}
For weak intrinsic variability, we write
\begin{align}
I_\mrm{S}(\mu,\b r;\f) = [I_0+\Delta I(\b r;\f)]f_\mrm{LD}(\mu) \, ,~
\Delta I\ll I_0 ~.
\end{align}
Keeping the leading term gives
\begin{align}
\label{Pgammagamma_surface_average}
\la P_{\gamma\gamma}\ra_\mrm{S}
\approx
\frac{3}{(3-u)\pi}
\int_{\mu>0}
f_\mrm{LD}(\mu)
P_{\gamma\gamma}(\b r,\f)
\,\mu\,d\Omega ~.
\end{align}
Equation \eqref{Pgammagamma_surface_average} is the phase-resolved axion template used in the statistical analysis. Its shape is fixed by the transverse-field distribution over the visible hemisphere, limb darkening, and the magnetic reconstruction at each rotational phase. The ordinary stellar variability is instead absorbed into $F_\mrm{bkg}$ and modeled phenomenologically in the next section.

For a broadband photometric observation, the measured flux is not the monochromatic flux at a single photon energy, but a response-weighted integral over the instrumental bandpass. The TESS response covers approximately $\lambda\simeq600$--$1000\,\mrm{nm}$, corresponding to photon energies
$\omega\simeq 1.2$--$2.1\,\eV$. To compare with observations, in principle, \eqref{Fobs} should be averaged over the instrumental response. In the coherent regime \eqref{light_axion_mass}, however, the leading conversion probability depends on the magnetic field integral, while the photon energy enters only through the subleading oscillatory phase. We therefore evaluate \eqref{Pagamma} at the effective photon energy $\omega=1.5498\,\eV$. This approximation preserves the relevant phase dependence for the low-mass axions constrained in this work and keeps the photometric template tied directly to the reconstructed MWD magnetic field.

\section{Magnetic field models}
\label{sec:magnetic_photometric_modeling}
The observed flux \eqref{Fobs1} requires two distinct inputs. The first is the magnetic geometry, which fixes the transverse-field integral in \eqref{Pagamma} for each visible surface element and rotation phase. The second is the intrinsic stellar light curve, which contains ordinary rotational variability unrelated to axion-photon conversion. In this section we specify the magnetic ingredients.

\subsection{Phase-resolved magnetic-field models}
The magnetic field of MWDs can be reconstructed from phase-resolved flux and circular-polarization spectra, following the Zeeman-tomography approach developed for MWDs \cite{Euchner:2002qv, Euchner:2005xk, Euchner:2006ku}. This involves parameter fitting for different magnetic field ansatz. 

One useful building block for magnetic fields is a dipole
\begin{align}
\label{B_dipole}
\b B(\b r') = \frac{B_0}{2}\l( \frac{R_\star}{r'} \r)^3 [ 3 \hat{\b r}' (\hat{\b m}\cdot\hat{\b r}') - \hat{\b m} ] ~,
\end{align}
where $\b r'$ refers to coordinates in the magnetic frame, $B_0$ is the signed polar field evaluated at $r'=R_\star$, and $\hat{\b m}$ is the dipole axis. A superposition of several off-centered, nonaligned dipoles can provide a compact phenomenological description of complex surface fields while retaining the correct dipolar falloff of each component away from the star.

Outside the stellar surface we approximate the magnetosphere as current-free, so that the magnetic field may be written in terms of a potential field, $\b B=-\nabla\Phi_B$. In a magnetic frame whose polar axis is aligned with $\hat{\b m}$, the scalar magnetic potential can be written in a multipole expansion,
\begin{align}
\label{Phi_B}
\Phi_\mrm{B} =& - R_\star \sum_{l=1}^{l_\mrm{max}} \sum_{m=0}^l \l( \frac{R_\star}{r'} \r)^{l+1} P_l^m(\cos\theta') \nonumber\\ &\times [ g_l^m \cos(m\j') + h_l^m \sin(m\j') ]  ~,
\end{align}
where $\theta'$ and $\j'$ are the polar and azimuthal angles with respect to the magnetic axis $\hat{\b m}$, $g_l^m$ and $h_l^m$ are magnetic multipole coefficients, and $P_l^m(\cos\theta')$ is the associated Legendre polynomial. Here, the $l=1$ terms reproduce the dipolar field, while higher multipoles encode the nondipolar structure inferred from the phase-resolved spectropolarimetry. Truncating at $l_\mrm{max}$ gives $(l_\mrm{max}+1)^2-1$ independent coefficients. 

To evaluate a magnetic model at a given rotation phase, we work in a frame where the line of sight is the $z$ axis. Denoting $i$ as the inclination between the rotation axis and the line of sight, we write the rotation axis as $\hat{\b\Omega}=(0,\sin i,\cos i)^\tran$. For a magnetic component tilted by an angle $\beta$ relative to $\hat{\b\Omega}$, the phase-dependent magnetic axis is
\begin{align}
\hat{\b m}(\f)
=
\begin{pmatrix}
\sin\beta\sin(2\pi\f+\j_0) \\
\sin i\cos\beta-\cos i\sin\beta\cos(2\pi\f+\j_0) \\
\cos i\cos\beta+\sin i\sin\beta\cos(2\pi\f+\j_0)
\end{pmatrix} ~,
\end{align}
where $\j_0$ is the azimuthal angle at phase $\f=0$. If the magnetic model is off-centered, the displacement vector can be written as $\b r_\mrm{off} = x_\mrm{off}' \hat{\b x}' + y_\mrm{off}' \hat{\b y}' + z_\mrm{off}' \hat{\b m}$, where we define a set of orthonormal vectors tied to the magnetic axis $\hat{\b y}' = (\hat{\b\Omega} \times \hat{\b m})/\sin\beta$ and $\hat{\b x}' = \hat{\b y}'\times\hat{\b m}$.

\subsection{PG 1015+014 as a benchmark target}
\label{sec:pg1015_014}
PG 1015+014 is a useful benchmark for a phase-resolved photometric axion search because of two reasons. First, the star is a rapidly rotating MWD, with a period of $98.7\,\mrm{min}$, so its magnetic topology is sampled many times in the TESS light curve. Second, its surface magnetic field has been reconstructed from rotation-phase-resolved optical flux and circular-polarization spectra obtained with FORS1 at the ESO VLT \cite{Euchner:2006ku}. These data show that the surface field strength is concentrated around $70$--$80\,\mrm{MG}$, with significant phase-dependent contributions to Zeeman features over roughly $50$--$90\,\mrm{MG}$. The combination of strong fields, rapid rotation, and an available magnetic reconstruction makes PG 1015+014 a controlled target for testing whether precision photometry can distinguish axion-induced phase structure from intrinsic stellar variability.

For the magnetic field, we use the two parametrizations that provide comparably good fits to the phase-resolved spectra in \cite{Euchner:2006ku}. The first is a superposition of three individually tilted and off-centered dipoles. In this model, the inclination of the rotation axis to the line of sight is $i\simeq 23\deg$. The three dipoles have signed polar field strengths $B_0=(-40,92,-38)\mrm{MG}$, tilt angles $\beta=(44\deg,63\deg,63\deg)$, and azimuthal angles $\j_0=(339\deg,276\deg,134\deg)$. Their displacement vectors $(x_\mrm{off}',y_\mrm{off}',z_\mrm{off}')$ in the corresponding magnetic frames are $(0.04, 0.35, 0.33) R_\star$, $(-0.012, -0.136, -0.28) R_\star$, and $(0.27, 0.080, 0.21) R_\star$ We evaluate each component with \eqref{B_dipole} and sum the fields before computing $B_\mrm{T}$ along each photon ray.

The second reconstruction is a truncated multipole expansion through $l_\mrm{max}=4$, with inclination $i\simeq 47\deg$, magnetic-axis tilt $\beta\simeq22\deg$, and azimuthal angle $\j_0=191\deg$. In using the published multipole coefficients, one must keep track of the convention for the associated Legendre functions. Directly inserting the tabulated coefficients of \cite{Euchner:2006ku} into \eqref{Phi_B} with unnormalized $P_l^m$ gives unphysically large surface fields, as noted in \cite{Tian:2025lkp}. We find that the published surface-field maps in \cite{Euchner:2006ku} are reproduced when the coefficients are interpreted with Schmidt semi-normalized associated Legendre functions,
\begin{align}
P_{l,\mrm{S}}^m(\cos\theta')
=
\left[
(2-\delta_{m0})\frac{(l-m)!}{(l+m)!}
\right]^{1/2}
P_l^m(\cos\theta') ~.
\end{align}
Equivalently, if \eqref{Phi_B} is written with unnormalized $P_l^m$, the coefficients must be rescaled by the normalization factor above. The coefficients in this convention are listed in table \ref{tab:multipole_coefficients}.

\begin{table}
\begin{tabular}{cccccc}
\hline
\hline
$\quad ~m~\quad$ & $\quad~~~\quad$ & $\quad l=1\quad$  & $\quad~2~\quad$ & $\quad~3~\quad$ & $\quad~4~\quad$ \\
\hline
0 & $g_l^m$ & $3.0$ & $0.6$ & $9.0$ & $2.4$ \\
1 & $g_l^m$ & $-12.5$ & $11.0$ & $-0.41$ & $1.5$ \\
 & $h_l^m$ & $-28.2$ & $4.1$ & $-6.37$ & $3.48$ \\
2 & $g_l^m$ & -- & $-5.66$ & $1.50$ & $-0.760$ \\
 & $h_l^m$ & -- & $-4.50$ & $-0.31$ & $0.48$ \\
3 & $g_l^m$ & -- & -- & $0.22$ & $-0.034$ \\
 & $h_l^m$ & -- & -- & $0.785$ & $-0.14$ \\
4 & $g_l^m$ & -- & -- & -- & $-0.033$ \\
 & $h_l^m$ & -- & -- & -- & $-0.062$ \\
\hline
\end{tabular}
\caption{Multipole coefficients used in \eqref{Phi_B} for the PG 1015+014 reconstruction. The coefficients are in units of $\mrm{MG}$ and correspond to the unnormalized associated Legendre convention after applying the Schmidt-normalization rescaling described in the text.}
\label{tab:multipole_coefficients}
\end{table}

\begin{figure}
\centering
\includegraphics[width=0.8\linewidth]{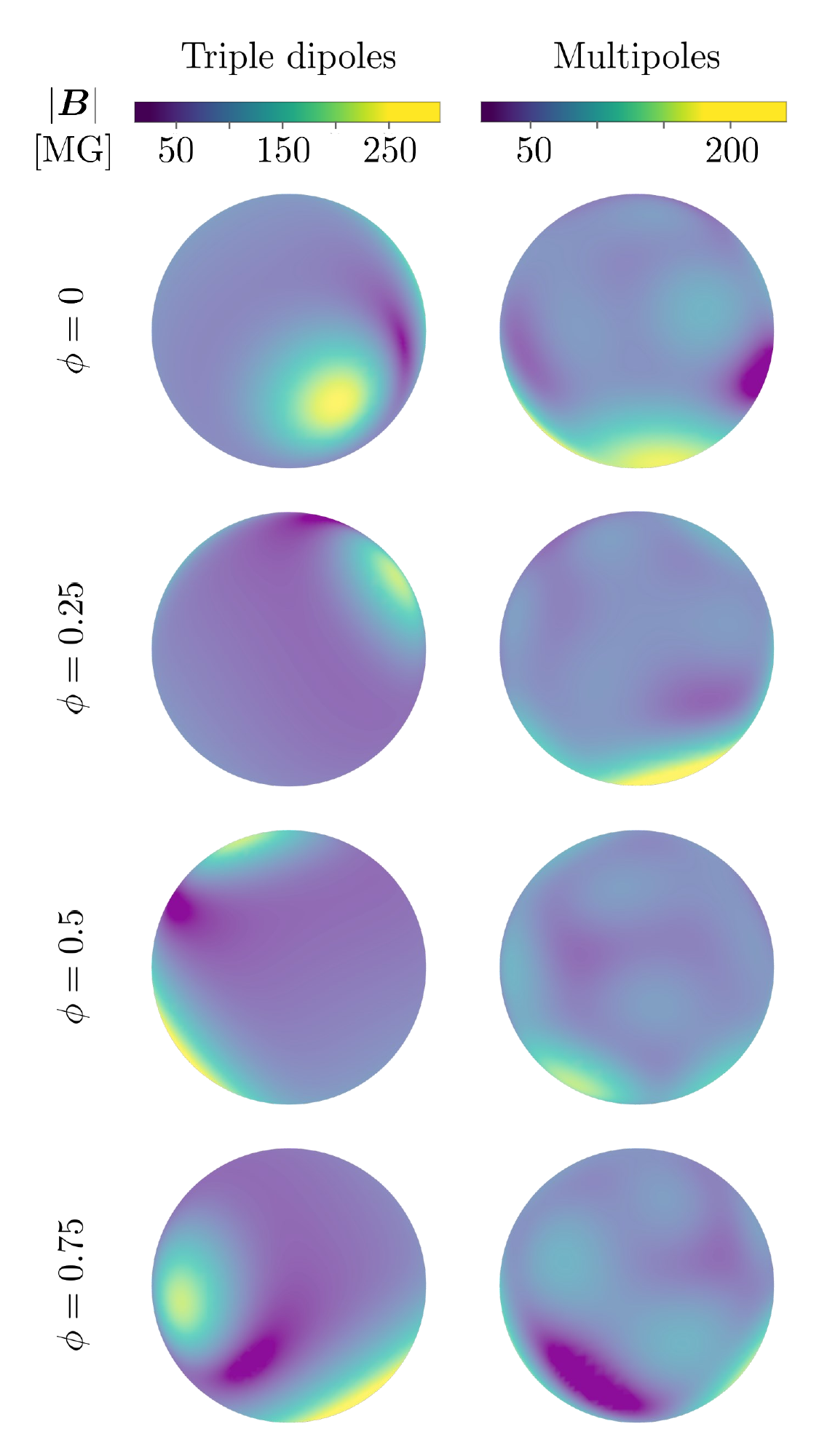}
\caption{Absolute value of the surface magnetic field on the visible hemisphere for the three off-centered-dipole reconstruction (left column) and the truncated multipole reconstruction with $l_\mrm{max}=4$ (right column). Rows show representative rotation phases $\phi=0$, $0.25$, $0.5$, and $0.75$.}
\label{fig:surfacefields}
\end{figure}

Figure \ref{fig:surfacefields} compares the visible-hemisphere surface-field distributions for the triple-dipole and multipole reconstructions at representative phases. Both reproduce the main phase-resolved spectropolarimetric features, while simpler parametrizations considered in \cite{Euchner:2006ku} do not fit all phases simultaneously. We therefore use the triple-dipole reconstruction for the main analysis and repeat the inference with the multipole reconstruction as a magnetic-model cross-check. 

We do not find reported values of the radius of PG 1015+014. Given its mass $M_\star=0.91\MSun$ \cite{2024arXiv240115158H}, we estimate the radius as $R_\star = 0.009R_\odot$ using the cold white dwarf mass-radius relation for $\mu_e=2$ \cite{1972ApJ...175..417N}.\footnote{In \cite{Tian:2025lkp}, the mass and radius of PG 1015+014 are assumed to be $M_\star=0.13 M_\odot$ and $R_\star=0.0247 R_\odot$. Such a large radius would overestimate the impact of axion-photon conversion as can be seen from \eqref{Pagamma}.} The limb-darkening coefficient is fixed to $u=0.47$, following the spectral modeling of PG 1015+014 \cite{Euchner:2006ku}.

\section{TESS data and phase-folded light curve}
\label{sec:tess}
We use optical photometry of PG 1015+014 from the Transiting Exoplanet Survey Satellite (TESS) \cite{2015JATIS...1a4003R}. TESS observes over a broad red-optical band, approximately $\lambda\simeq600$--$1000\,\mrm{nm}$, matching the photon-energy range used in the axion-conversion calculation. The available observations cover six sectors, S35, S45, S46, S62, S72, and S89, with 20$\s$ cadence data spanning BJD $2459255.829$ to $2460745.469$. We retrieve the light curves from the Mikulski Archive for Space Telescopes (MAST) \cite{MAST_TESS_PG1015}. Since the axion template is a small phase-dependent distortion of the stellar light curve, the data reduction must preserve rotational variability near the $98.7\,\mrm{min}$ period while removing long-timescale instrumental trends.

We work with the Pre-search Data Conditioning Simple Aperture Photometry (PDCSAP) flux. The PDCSAP pipeline corrects common instrumental systematics, including scattered light, thermal effects, and spacecraft trends, while aiming to preserve astrophysical variability \cite{Stumpe:2012xg, Smith:2012xh}. The uncorrected SAP flux retains larger sector-dependent systematics and is therefore less suitable for constructing a phase-folded light curve at the precision required here. We normalize each sector before combining the data, so that the subsequent fit is sensitive to the rotational phase structure rather than to sector-to-sector flux offsets.

We further clean the 20$\s$ cadence light curves with the \texttt{Lightkurve} package \cite{2018ascl.soft12013L}. Isolated outliers are removed by iterative $5\sigma$ clipping. This threshold is conservative enough to avoid sculpting the rotational modulation while eliminating sharp single-cadence excursions, which are typically associated with cosmic-ray hits or detector artifacts. After this step, we remove long-timescale residual trends with a sliding quadratic filter using a window length of 8000 cadences, corresponding to approximately $44\,\hr$. This window is much longer than the rotation period and therefore removes sector-scale instrumental drifts without fitting away the phase structure at the stellar spin frequency and its first few harmonics.

The cleaned and normalized data are folded on the stellar rotation period,
\begin{align}
\label{phasefold}
\f_i
=
\mrm{mod}
\left(
\frac{t_i-t_0}{P},1
\right) ~,
\end{align}
where $P$ is the rotation period inferred from the Lomb-Scargle periodogram \cite{Virtanen:2019joe} using the python module \texttt{astropy}. We choose the reference epoch $t_0$ to be the time of maximum flux in the densely sampled light curve. This convention fixes only the photometric phase origin; the relative phase between the photometric light curve and the magnetic reconstruction is still allowed to vary through $\f_{a\gamma}$ in \eqref{Fobs2}. Because the TESS sectors are separated over a baseline of about four years, phase folding combines many disjoint rotation cycles into a single high signal-to-noise light curve while retaining the phase dependence relevant for axion-photon conversion.

For the Bayesian analysis, we bin the folded light curve in phase. In each bin $j$, the weighted mean flux and formal uncertainty are
\begin{align}
\bar F_j = \frac{\sum_{i\in j} w_i F_i}{\sum_{i\in j} w_i} \sep
\sigma_{\bar F_j} = \left(\sum_{i\in j}w_i\right)^{-1/2} ~,
\end{align}
where the summation runs over all data points in the $j$th bin and $w_i = \sigma_i^{-2}$ is the inverse of the squared uncertainty of each data point. The bin width is chosen to provide adequate phase resolution for the rotational modulation while ensuring sufficient data points per bin for statistically meaningful averages. We have checked that the scatter error of each bin is less than or comparable to the formal uncertainty to render a conservative analysis.

\begin{figure}
\centering
\includegraphics[width=1\linewidth]{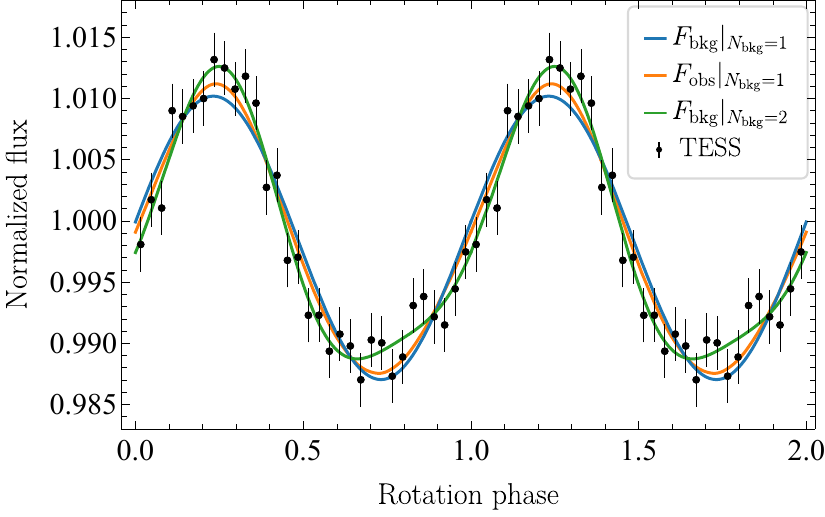}
\caption{Phase-folded TESS light curve of PG 1015+014. Data points are extracted after sector normalization, outlier rejection, long-timescale detrending, and inverse-variance binning. The curves show representative best-fit models: a sinusoidal background with $N_\mrm{bkg}=1$, a two-harmonic background with $N_\mrm{bkg}=2$, and the full axion-plus-background fit for $N_\mrm{bkg}=1$.}
\label{fig:lightcurve}
\end{figure}

The resulting phase-folded light curve with $32$ bins is shown in figure \ref{fig:lightcurve}. The data exhibit a nearly periodic rotational modulation, but the shape is not forced to be sinusoidal. This is precisely the regime in which the background model of \eqref{Fbkg} is needed: the same folded data will be fit with $N_\mrm{bkg}=1$ and $N_\mrm{bkg}=2$ to test whether any apparent axion preference survives the inclusion of the leading higher harmonic.

\section{Intrinsic photometric variability and background light curves}
\label{sec:background_model}
The phase-resolved axion template must be compared against ordinary rotational variability of the star. Such variability is common in MWDs and need not be sinusoidal. The physical origin can differ from object to object. For instance, changes in the visible magnetic-field distribution can move Zeeman-split spectral features through the bandpass, magnetic dichroism can modulate the continuum opacity in high-field objects, and surface-brightness inhomogeneities or magnetic spots can generate rotational flux modulation even at lower field strengths \cite{2013ApJ...773...47B, Maxted:2000wp, Valyavin:2011rb, Maoz:2014upa, 2024arXiv240115158H}.

Rather than modeling the atmosphere, opacity, and surface temperature distribution from first principles, we describe the intrinsic stellar light curve phenomenologically. For a phase-folded light curve, the background flux is written as a truncated Fourier series,
\begin{align}
\label{Fbkg}
F_\mrm{bkg}(\f)
=
\theta_0
+
\sum_{n=1}^{N_\mrm{bkg}}
\left[
\theta_{2n-1}\sin(2\pi n\f)
+
\theta_{2n}\cos(2\pi n\f)
\right] ~,
\end{align}
where the coefficients $\theta_i$ are nuisance parameters. The truncation order $N_\mrm{bkg}$ controls the amount of non-axion phase structure allowed in the fit. The choice $N_\mrm{bkg}=1$ corresponds to a purely sinusoidal stellar background, which is often an effective empirical description of MWD rotational modulation \cite{2013ApJ...773...47B} and was adopted in the recent PG 1015+014 photometric axion analysis \cite{Tian:2025lkp}. However, nondipolar magnetic topology, phase-dependent Zeeman opacity, localized surface structure, and limb-darkening effects can naturally generate higher harmonics.

Following the flux relation \eqref{Fobs1}, the full model for the phase-folded flux is then 
\begin{align}
\label{Fobs2}
F_\mrm{obs}(\f) = F_\mrm{bkg}(\f; \b\theta) \la P_{\gamma\gamma}\ra_\mrm{S} \left(\f+\f_{a\gamma};m_a,g_{a\gamma}\right) ~.
\end{align}
Here $\f_{a\gamma}$ is a relative phase offset between the photometric reference phase and the magnetic-field reconstruction. This parameter accounts for uncertainties in matching the phase convention of the TESS light curve to that of the spectropolarimetric magnetic model. Since $P_{a\gamma}\ll 1$ in the parameter region of interest, the axion contribution is a small phase-dependent distortion of the stellar background.

The role of the Fourier background is therefore diagnostic as well as statistical. If $N_\mrm{bkg}$ is too small, ordinary nonsinusoidal stellar structure can be misidentified as an axion contribution. If $N_\mrm{bkg}$ is too large, the background can absorb phase structure that would otherwise indicate axion-photon conversion. We therefore compare the sinusoidal model, $N_\mrm{bkg}=1$, with higher-harmonic extensions, such as $N_\mrm{bkg}=2$. As we will see in section \ref{sec:bayesian_constraints}, for the TESS data of PG 1015+014, this comparison is the central background-degeneracy test. An apparent axion preference under a sinusoidal background should not be interpreted as evidence for axions unless it provides a better fit than the background model with the inclusion of higher harmonics.

\section{Bayesian inference and projected sensitivity}
\label{sec:bayesian_constraints}
We now use the phase-folded TESS light curve to constrain the axion-photon coupling. We adopt a Gaussian likelihood for the binned light curve,
\begin{equation}
\label{eq:likelihood}
\log P(\text{data}|\text{model}) = -\frac{1}{2} \sum_j \left[ \frac{ \bar F_j-F_\mrm{obs}(\f_j;\b\Theta) }{ \sigma_{\bar F_j} } \right]^2 ~,
\end{equation}
where the sum is over all binned data points. The sampled model parameters $\b\Theta$ include the Fourier coefficients $\b\theta$ and the axion-related parameters $\log_{10}(m_a/\eV)$, $\log_{10} (g_{a\gamma}/\GeV^{-1})$, and $\f_{a\gamma}$. We use a relatively wide Gaussian prior on the mean flux, $\theta_0\sim N(1,0.01^2)$, broad uniform priors on the remaining Fourier coefficients, and flat priors for axion parameters $\log_{10}\left(m_a/\eV\right)\in[-7.5,-6]$, $\log_{10} \left(g_{a\gamma} / \GeV^{-1}\right) \in [-11.2,-10.2]$, and $\f_{a\gamma}\in[-0.5,0.5]$. The posterior is sampled with a Metropolis-Hastings Markov chain. For each magnetic-field and background model, we run $16$ or more independent chains with $10^5$ steps each and discard the first $2000$ steps as burn-in. The proposal covariance is tuned to give an acceptance fraction of $\sim 10 \%$--$30\%$. We verify convergence by comparing the marginalized posteriors from independent chains, checking that the inferred upper limit on $g_{a\gamma}$ is stable when the burn-in fraction and chain length are varied. We have also verified that the following analysis is not significantly affected when the data are binned into either $24$ or $48$ bins. Unless otherwise stated, the Bayesian results below use the triple-dipole magnetic model described in section \ref{sec:pg1015_014}. The results of the multipole reconstruction are provided in appendix \ref{app:multipole}, which give compatible constraints and are used as a modeling cross-check.

\subsection{Posterior distribution for the sinusoidal background}
\begin{figure}
\centering
\includegraphics[trim=0 120 0 120, clip, width=\linewidth]{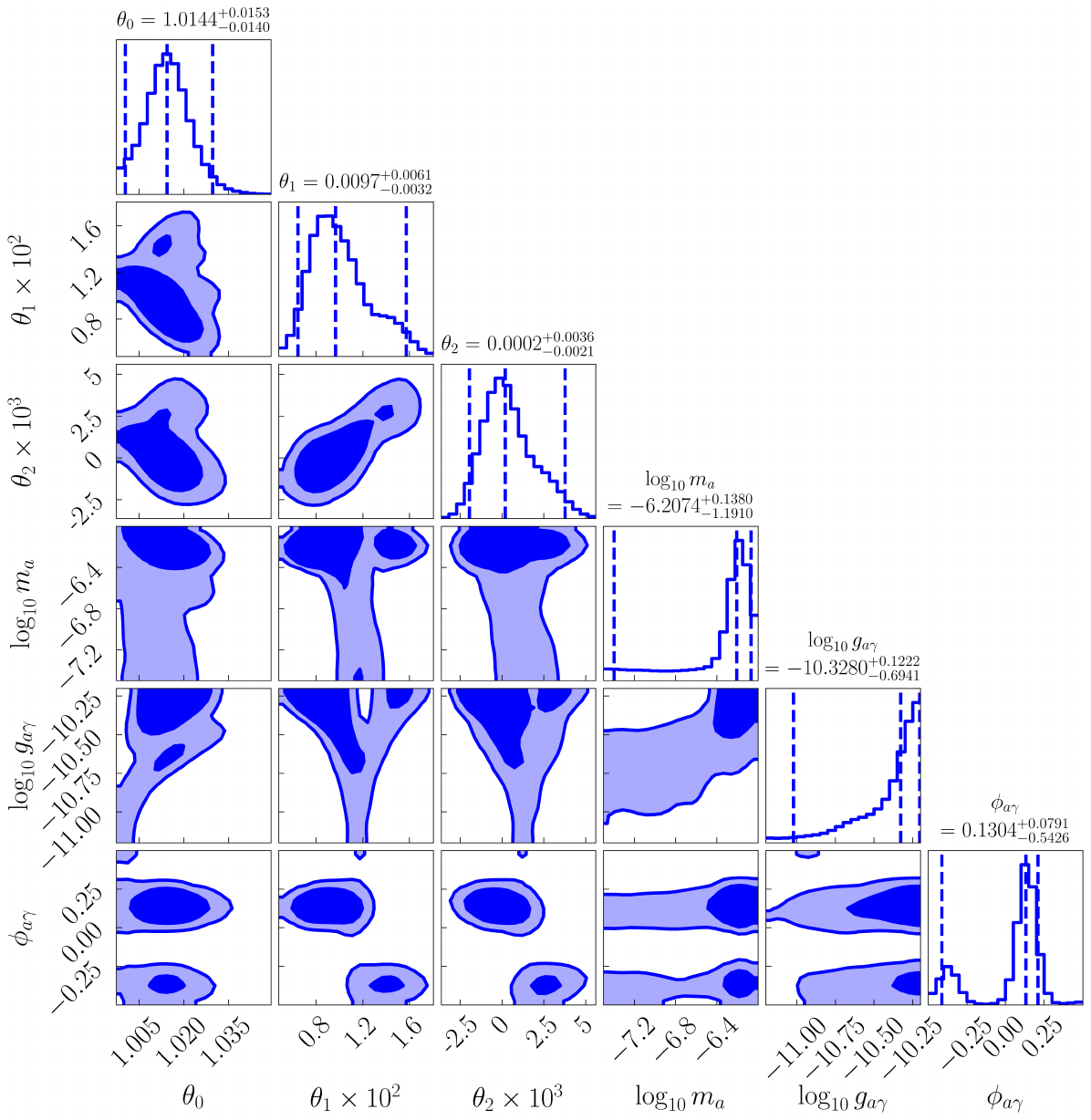}
\caption{Posterior distribution for the triple-dipole magnetic model with a sinusoidal intrinsic background, $N_\mrm{bkg}=1$.  The fit develops apparent support for nonzero $g_{a\gamma}$, but this preference is driven by residual stellar variability that is not captured by a single harmonic.  Allowing a second harmonic in the background removes this apparent axion preference, see figure \ref{fig:cornernbkg23dipoles}.}
\label{fig:cornernbkg13dipoles}
\end{figure}

A sinusoidal background, $N_\mrm{bkg}=1$, is a natural first diagnostic because it captures the leading rotational modulation and is often adequate for MWD light curves at current precision. The posterior distribution is shown in figure \ref{fig:cornernbkg13dipoles}, where the shaded regions represent $68\%$ and $95\%$ confidence intervals and the vertical lines in histograms refer to $(0.025, 0.5, 0.975)$-th quantiles. With $N_\mrm{bkg}=1$, the posterior develops support away from zero coupling, with a representative region near $m_a\sim 10^{-6.2}\,\eV$ and $g_{a\gamma}\sim 10^{-10.3}\,\GeV^{-1}$. However, this behavior should not be interpreted as evidence for axion-photon conversion. Rather, the axion template supplies phase structure that the purely sinusoidal stellar background cannot reproduce. In next subsection, we will see that the same residual structure is absorbed by a higher-harmonic background model without invoking axions.

This degeneracy is physically expected. MWDs can exhibit nonsinusoidal rotational modulation because the surface brightness, magnetic-field distribution, atmospheric opacity, and viewing geometry need not be dominated by a single harmonic. Since axion-photon conversion also produces a phase-dependent attenuation fixed by the magnetic geometry, an overly rigid background model can convert ordinary stellar variability into an apparent axion preference.  The sinusoidal-background posterior therefore provides a useful stress test of the analysis, but not a conservative constraint.

\subsection{Posterior distribution for the two-harmonic background}
\begin{figure*}
\centering
\includegraphics[trim=0 120 0 120, clip, width=0.8\linewidth]{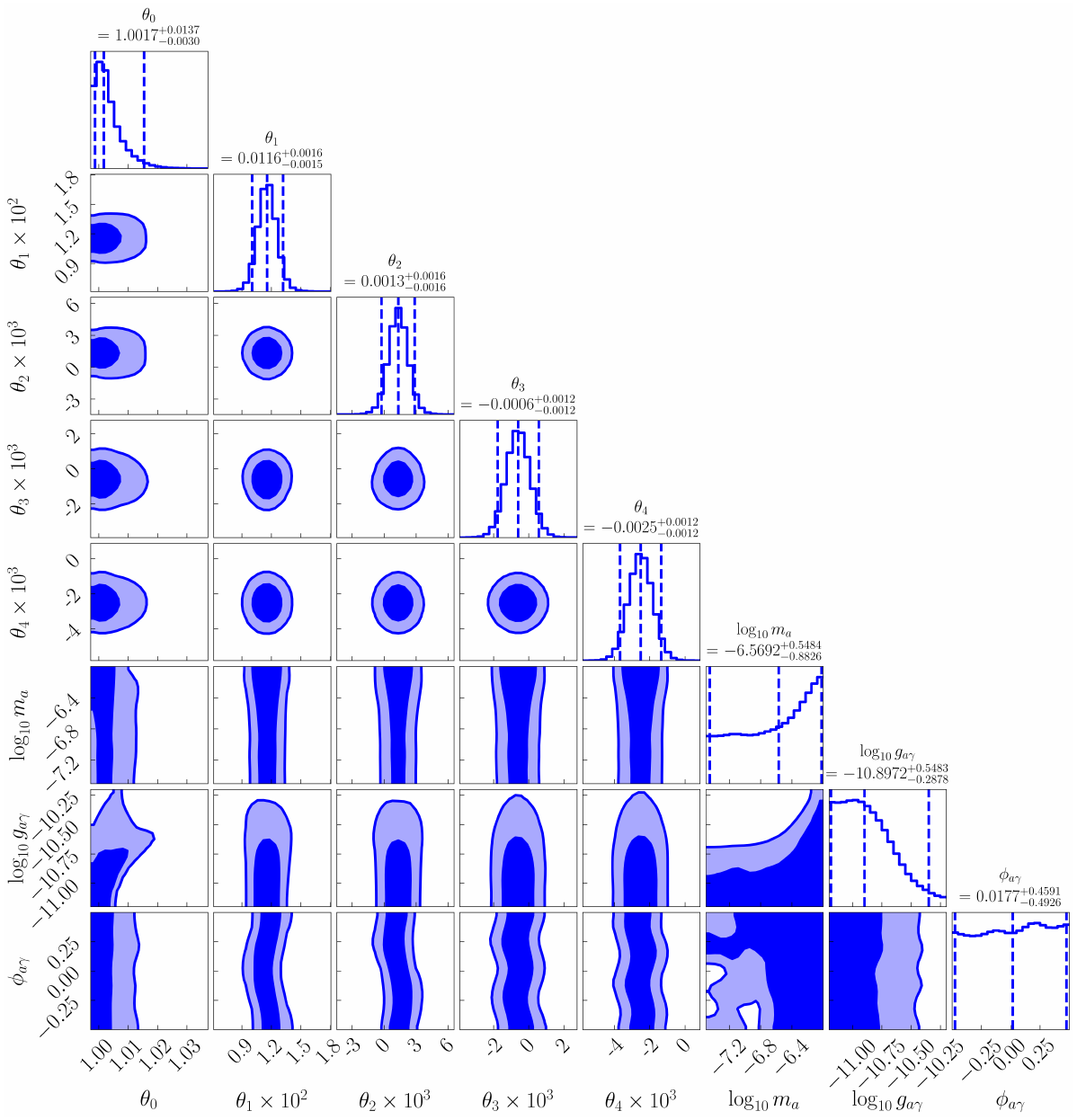}
\caption{Posterior distribution for the triple-dipole magnetic model with the baseline two-harmonic intrinsic background, $N_\mrm{bkg}=2$.  The apparent preference for nonzero coupling found in the sinusoidal-background fit in figure \ref{fig:cornernbkg13dipoles} is absorbed by the additional stellar harmonic. The resulting marginalized posterior yields the baseline $95\%$ upper limit on axion-photon coupling, given by \eqref{axion_limit}.}
\label{fig:cornernbkg23dipoles}
\end{figure*}

We therefore take $N_\mrm{bkg}=2$ as the baseline model for setting a conservative limit. This model includes the leading higher harmonic of the intrinsic rotational variability while retaining enough structure to avoid fitting arbitrary phase-dependent features. The posterior distribution is shown in figure \ref{fig:cornernbkg23dipoles}. As we can see, the posterior no longer shows a preference for nonzero axion-photon conversion. The coupling posterior is instead concentrated toward the smallest allowed values, and the data constrain any additional axion-induced attenuation to be subdominant to the observed stellar modulation. For axion masses in the coherent regime of equation \eqref{light_axion_mass}, the conversion probability is nearly independent of $m_a$, and the constraint is correspondingly flat in mass. We quote the marginalized $95\%$ upper limit
\begin{equation}
\label{axion_limit}
g_{a\gamma} < 2\times10^{-11}\,\GeV^{-1} \quad\text{for}\quad
m_a\lesssim 2\times10^{-7}\,\eV ~.
\end{equation}
The multipole magnetic reconstruction gives a similar limit, despite its different surface-field morphology (Appendix \ref{app:multipole}). This agreement indicates that the result is not driven by a single idealized magnetic geometry.

\subsection{Model selection and saturation test}
\label{sat_test}
To quantitatively assess whether the apparent axion preference under the sinusoidal background is statistically justified, we compare competing models using information criteria. We compare two models:
\begin{itemize}
\item \textbf{Model A}: \(N_{\mathrm{bkg}}=1\) background + axion. This model has \(k=6\) parameters: 3 Fourier coefficients (\(\theta_0,\theta_1,\theta_2\)) plus 3 axion parameters (\(\log_{10}m_a\), \(\log_{10}g_{a\gamma}\), \(\phi_{a\gamma}\)).
\item \textbf{Model B}: \(N_{\mathrm{bkg}}=2\) pure background. This model has \(k=5\) parameters: 5 Fourier coefficients (\(\theta_0,\theta_1,\theta_2,\theta_3,\theta_4\)).
\end{itemize}
Using the best-fit \(\chi^2\) values from the MCMC chains, we find \(\chi^2_A = \chi^2_B + 9.94\). The Akaike Information Criterion (AIC)~\cite{Akaike:1974vps} and Bayesian Information Criterion (BIC)~\cite{Schwarz:1978tpv} are defined as
\begin{equation}
\mathrm{AIC} = \chi^2 + 2k, \qquad
\mathrm{BIC} = \chi^2 + k\ln N,
\end{equation}
where \(N\) is the number of binned data points (\(N=32\)). This gives
\begin{align}
\Delta\mathrm{AIC} &= \mathrm{AIC}_A - \mathrm{AIC}_B = 11.94, \\
\Delta\mathrm{BIC} &= \mathrm{BIC}_A - \mathrm{BIC}_B = 13.52.
\end{align}
Both exceed the conventional threshold of 10, indicating strong evidence against Model A and in favor of Model B \cite{Kass:1995loi}. The apparent axion preference under the sinusoidal background is thus a statistical artifact due to an overly restrictive background model.

To ensure that our baseline \(N_{\mathrm{bkg}}=2\) is not an arbitrary truncation, we have extended the background model to \(N_{\mathrm{bkg}}=3\) (with 7 Fourier coefficients). We find that the difference between the best-fit $\chi^2$ values of the $N_\mrm{bkg}=3$ background-only model and the $N_\mrm{bkg}=2$ one is $\Delta\chi^2=-0.10$, implying that the two models are essentially indistinguishable within the current data. We conclude that including additional harmonics would not appreciably change the inferred limit.

\subsection{Projected reach for other magnetic white dwarfs}
\label{other_MWD}
The analysis of PG~1015+014 shows that a phase-resolved magnetic model and a flexible stellar background are both needed before a photometric modulation can be interpreted as an axion constraint.  Nevertheless, the expected sensitivity of other MWDs can be estimated without performing a full reconstruction. The purpose of this estimate is not to assign object-specific limits to individual stars, but to identify systems for which existing TESS light curves and strong magnetic fields make phase-resolved follow-up especially valuable.

The scaling follows directly from the coherent-regime estimate in section \ref{sec:axion_photon_conversion}. Let $B_0$ denote a characteristic surface magnetic field and let $\sigma_\mrm{eff}$ be the normalized uncertainty of the phase-binned light curve after folding at the stellar rotation period. The surface-averaged axion signal is smaller than the ray-level estimate because different surface elements have different transverse-field integrals and because only the phase-dependent part of the attenuation can be distinguished from the intrinsic background. We parametrize this reduction by $\la P_{a\gamma}\ra_\mrm{S} = x\,P_{a\gamma}$, where $P_{a\gamma}$ is obtained from the estimate below Eq.~\eqref{light_axion_mass}, and $x$ is an effective geometric and surface-averaging factor. Requiring the corresponding fractional flux modulation, $\la P_{a\gamma}\ra_\mrm{S}^\mrm{var}/2$, to be smaller than $3\sigma_\mrm{eff}$ gives the target-ranking estimate
\begin{align}
\label{gagamma_constraint_expectation}
g_{a\gamma}^\mrm{lim}
&\sim
2\times 10^{-11}\,\GeV^{-1}
\left( \frac{\sigma_\mrm{eff}}{0.002} \right)^{1/2}
\left( \frac{0.05}{x} \right)^{1/2}
\nonumber\\
&\quad\times
\left( \frac{200\,\mrm{MG}}{B_0} \right)
\left( \frac{0.009\RSun}{R_\star} \right) ,
\end{align}
valid for $m_a\lesssim 2\times10^{-7}\,\eV$ for the fiducial white-dwarf radius. The numerical normalization is chosen to match the coupling scale reached in the full PG~1015+014 analysis.  Equation \eqref{gagamma_constraint_expectation} should therefore be viewed as an analytic sensitivity estimate, not as a substitute for the Bayesian treatment used above. In particular, the value of $x$ is object dependent and can only be computed reliably once the magnetic topology, inclination, limb darkening, and background light curve are modeled.

\begin{figure}
\centering
\includegraphics[width=\linewidth]{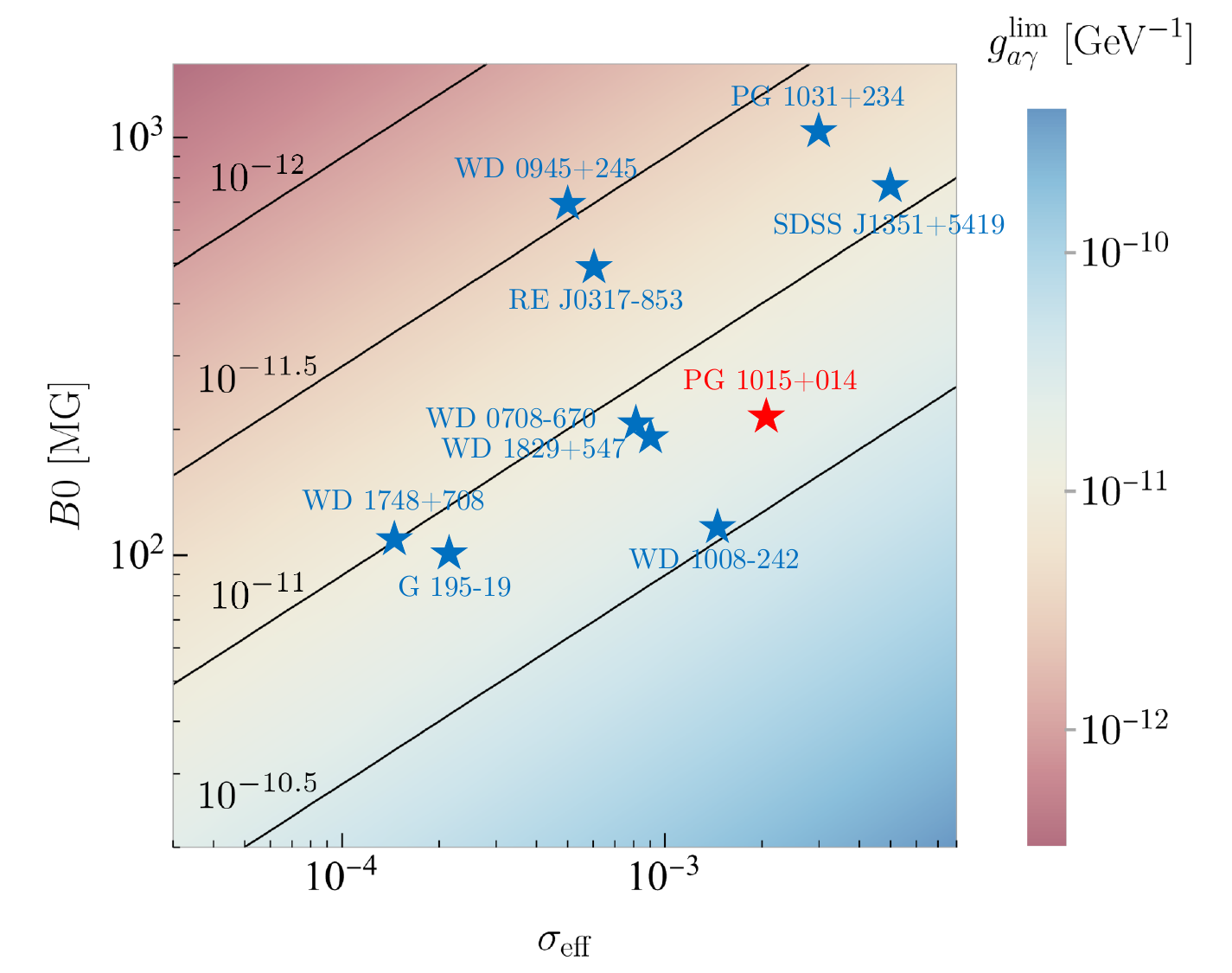}
\caption{Projected reach for the axion-photon coupling from existing TESS photometry of MWDs, estimated with equation \eqref{gagamma_constraint_expectation}. The red star shows the PG~1015+014 bound obtained in this work, while the blue stars show target-ranking projections for other MWDs. Here $B_0$ is the characteristic magnetic field, $\sigma_\mrm{eff}$ is the normalized uncertainty of the phase-binned light curve, and the projection fixes $R_\star=0.009\RSun$ and $x=0.05$ for all targets.  The diagonal guide lines indicate representative values of $g_{a\gamma}^\mrm{lim}$. }
\label{fig:projection}
\end{figure}

Figure \ref{fig:projection} applies this estimate to PG~1015+014 and to nine additional TESS MWD targets. For the projected targets, we estimate $\sigma_\mrm{eff}$ from the formal uncertainty of a $32$-bin phase-folded TESS light curve using the available $20\s$ cadence data, and we take the characteristic magnetic fields from the MWD compilations of \cite{Ferrario:2015oda, 2024A&A...692A.174B}. For a uniform comparison, the projection fixes $R_\star=0.009\RSun$ and $x=0.05$. These choices isolate the two measured quantities that most directly control the reach, namely the photometric precision and the magnetic-field scale.  They also make clear how the estimate should be rescaled once object-specific radii or phase-resolved magnetic maps become available.

The projected targets should therefore be interpreted as priorities for detailed modeling rather than as sources with established axion limits. Several of them combine stronger characteristic fields or smaller binned uncertainties than PG~1015+014, suggesting that a dedicated analysis with spectropolarimetric magnetic reconstruction could improve substantially on the benchmark constraint in \eqref{axion_limit}.  Conversely, a target with excellent TESS precision may still yield a weak axion constraint if its phase-dependent magnetic geometry produces a small value of $x$, or if intrinsic stellar harmonics closely mimic the axion template. The main implication of figure \ref{fig:projection} is thus operational: existing TESS photometry already contains a set of MWDs for which magnetic tomography and background-marginalized phase fitting could extend this time-domain axion search beyond a single benchmark object.

\section{Summary and outlook}
\label{sec:conclusions}
We have developed a phase-resolved photometric framework for testing axion-photon conversion with rotating MWDs while explicitly marginalizing over intrinsic stellar harmonics.  The central ingredient is that the magnetic topology fixes a rotation-dependent photon survival probability, while the background stellar light curve is treated as a phenomenological Fourier series. For a given magnetic reconstruction, we compute the surface-averaged survival probability over the visible stellar disk and fit the resulting axion template simultaneously with the intrinsic photometric variability. This formulation turns the main obstacle for photometric axion searches in MWDs — the degeneracy between axion-induced attenuation and stellar rotational structure — into a controlled background-modeling problem.

Applying this framework to TESS observations of PG~1015+014, we find that the interpretation depends crucially on the assumed stellar background. With a purely sinusoidal background, the posterior develops an apparent preference for nonzero axion-photon conversion, reproducing the qualitative behavior expected from a restrictive background model. This preference is not stable once the intrinsic light curve is allowed to contain the leading higher harmonic: the same phase structure is absorbed by the stellar background, and the axion posterior is driven toward small coupling. Indeed, model selection via AIC and BIC strongly favors the two-harmonic background over the axion-plus-sinusoidal model, with $\Delta\mathrm{AIC}=11.94$ and $\Delta\mathrm{BIC}=13.52$. Allowing a third harmonic in the background model produces no material change in the goodness of fit, indicating that the result has converged with respect to the tested background complexity.

Taking the two-harmonic background as a conservative baseline, and using the radius inferred from the white-dwarf mass-radius relation, we obtain $g_{a\gamma}<2\times10^{-11}\,\GeV^{-1}$ for $m_a\lesssim2\times10^{-7}\,\eV$ at $95\%$ confidence. The triple-dipole and multipole magnetic reconstructions give compatible constraints, indicating that this conclusion is not tied to a single parametrization of the PG~1015+014 surface field. The main distinction from the previous PG~1015+014 photometric work \cite{Tian:2025lkp} is therefore interpretive as well as technical. We replace the purely sinusoidal intrinsic light curve with a flexible Fourier background, reinterpret the sinusoidal-background axion preference as a degeneracy with intrinsic stellar harmonics, repeat the analysis with two independent magnetic reconstructions, and use a radius estimate tied to the measured white dwarf mass. These changes shift the emphasis from a single-object bound to the conditions under which a phase-resolved photometric search can separate an axion template from ordinary stellar variability.

We also derived an analytic sensitivity estimate for prioritizing other TESS magnetic white dwarfs. The projected reach scales as $g_{a\gamma}^\mrm{lim} \propto \sigma_\mrm{eff}^{1/2} x^{-1/2} B_0^{-1} R_\star^{-1}$, where $\sigma_\mrm{eff}$ is the phase-binned photometric uncertainty and $x$ encodes the suppression from surface averaging and magnetic geometry. Applying this scaling identifies targets for which strong fields and precise light curves make phase-resolved follow-up worthwhile. These projections are not limits on individual stars; they are target priorities for future analyses that include magnetic tomography, object-specific radii, limb darkening, and background marginalization.

Our work highlights a general caution for any time-domain search for exotic physics using stellar photometry: the intrinsic harmonic content of the star must be treated as a leading systematic, rather than absorbed into a rigid sinusoidal prior. The framework we have developed---combining a physically motivated template with a flexible background model and rigorous model selection---could be potentially adapted to other magnetized objects (e.g., neutron stars, active stars) and other time-domain probes (e.g., radio, X-rays). The key requirement is a sufficiently well-characterized model of the source variability, either from first principles or from empirical reconstruction.

Several improvements can sharpen this program.  Phase-resolved spectropolarimetry of the projected targets would turn the current scaling estimates into quantitative axion search. The accumulation of TESS data will further reduce $\sigma_\mrm{eff}$ and test whether the relevant stellar harmonics remain stable over long baselines. The ongoing and upcoming generation of wide-field time-domain facilities---including the Vera C. Rubin Observatory's Legacy Survey of Space and Time (LSST) \cite{LSST:2008ijt}, Nancy Grace Roman Space Telescope (WFIRST) \cite{Akeson:2019biv}, the China Space Station Telescope (CSST) \cite{CSST:2025ssq}, and the Ultraviolet Transient Astronomy Satellite (ULTRASAT) \cite{Shvartzvald:2023ofi}---will provide high-cadence, multiband photometry for orders of magnitude more MWDs, potentially reducing $\sigma_{\mathrm{eff}}$ and probing different axion mass windows via their different wavelength coverages. More detailed atmosphere and bandpass modeling could also connect the Fourier background to the underlying Zeeman opacity, surface brightness, and magnetic-field distribution. The broader path is clear: MWD photometry becomes a controlled axion search when precision light curves are paired with phase-resolved magnetic field models and a background model flexible enough to absorb ordinary stellar variability.

\begin{acknowledgments}
We would like to thank Dong Lai for helpful discussions. This research is supported by the program of the National Natural Science Foundation of China (grant Nos. 12473043, 12433013). HYZ acknowledges support from the Shanghai Magnolia Plan Pujiang Program 25PJA073 and CPSF General Program 2025M783412.
\end{acknowledgments}

\appendix
\section{Bayesian constraints for the magnetic multipole model}
\label{app:multipole}
In the main text we used the three off-centered-dipole reconstruction of PG~1015+014 as the baseline magnetic model.  Here we repeat the Bayesian analysis with the independent multipole reconstruction described in section \ref{sec:pg1015_014}.  The purpose of this appendix is to test whether the inference is tied to a particular magnetic geometry used to compute the phase-resolved axion template.  The likelihood, priors, TESS light curve, limb-darkening prescription, and Fourier background models are kept identical to those in section \ref{sec:bayesian_constraints}; only the magnetic field entering $\la P_{\gamma\gamma}\ra_\mrm{S}$ is changed.

\begin{figure}
\centering
\includegraphics[trim=0 120 0 120, clip, width=1\linewidth]{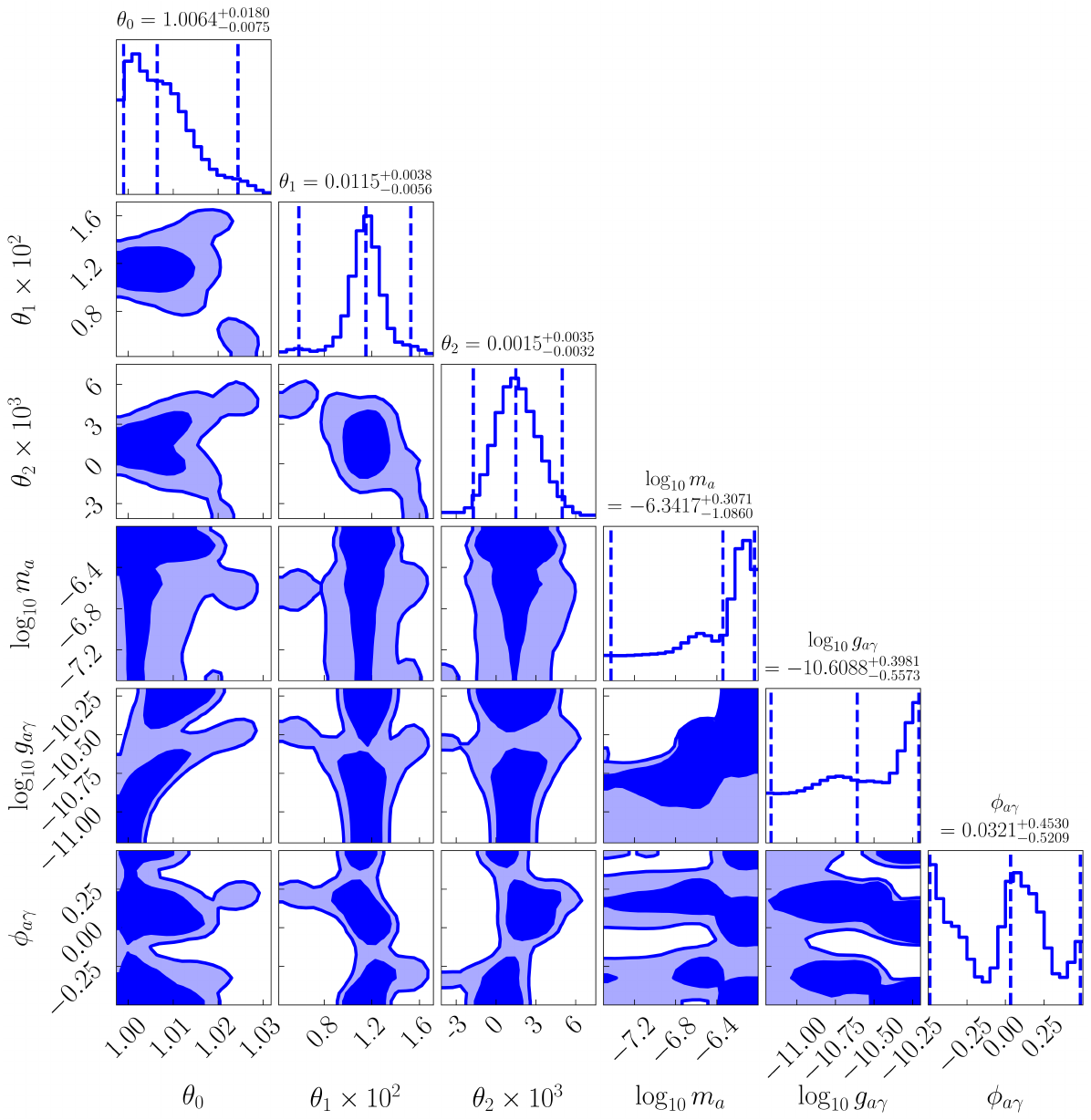}
\caption{Posterior distribution for the multipole magnetic reconstruction with a sinusoidal intrinsic background, $N_\mrm{bkg}=1$.  As in the triple-dipole analysis of figure \ref{fig:cornernbkg13dipoles}, the restricted background leaves residual phase structure that produces apparent support for nonzero $g_{a\gamma}$.  This preference is removed when the leading higher harmonic is included in the background model, see figure \ref{fig:cornernbkg2multipole}.}
\label{fig:cornernbkg1multipole}
\end{figure}

Figure \ref{fig:cornernbkg1multipole} shows the posterior for the multipole model with a sinusoidal intrinsic background, $N_\mrm{bkg}=1$.  As in the triple-dipole analysis, the restricted background leaves residual phase structure that can be partially fit by the axion template.  The posterior therefore develops apparent support for nonzero $g_{a\gamma}$.  This behavior is not a separate indication of axion-photon conversion.  Instead, it reinforces the conclusion of the main text: when the stellar background is forced to be purely sinusoidal, ordinary nonsinusoidal rotational variability can be misassigned to the magnetic axion template.

\begin{figure}
\centering
\includegraphics[trim=0 120 0 120, clip, width=1\linewidth]{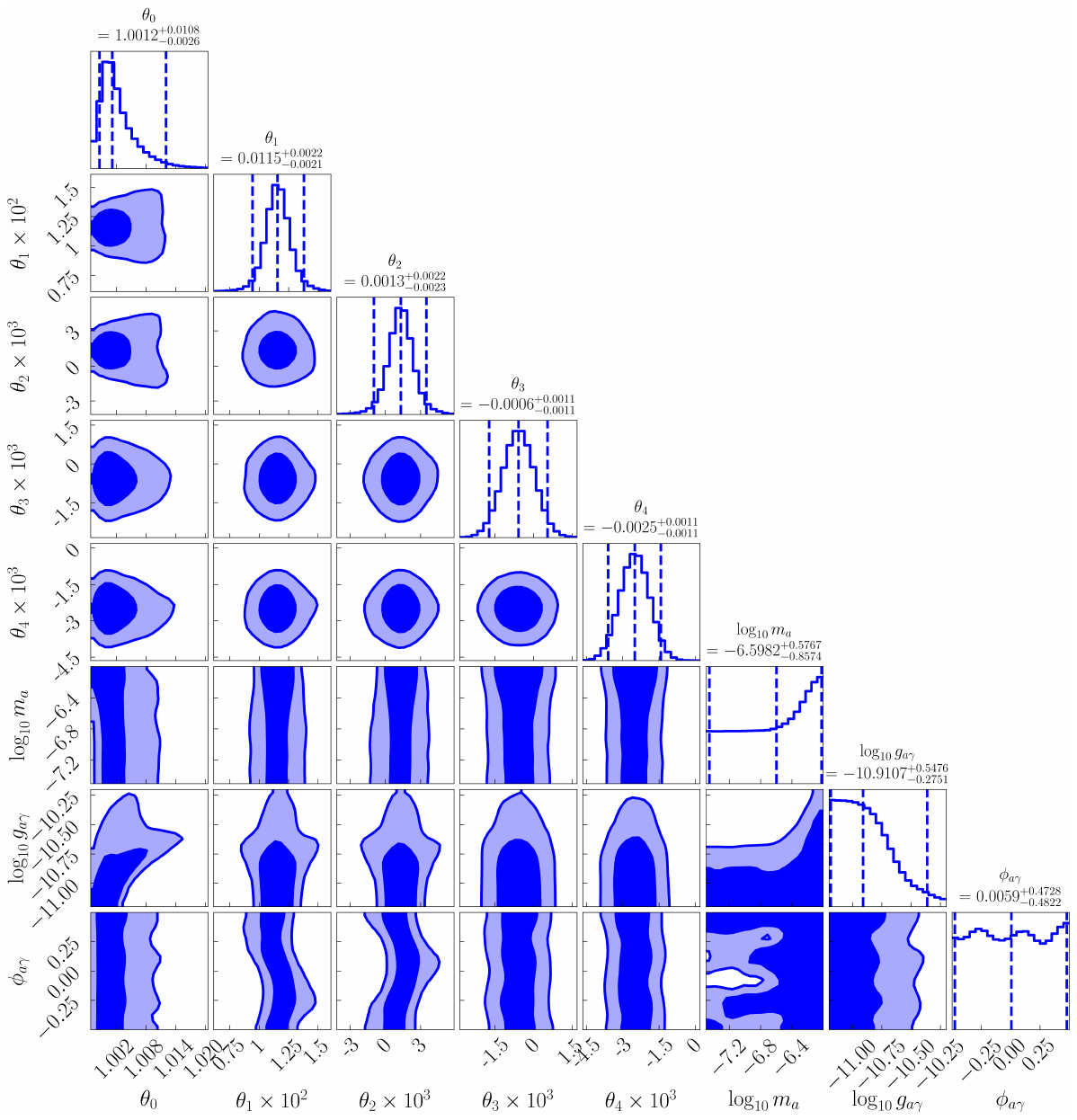}
\caption{Posterior distribution for the multipole magnetic reconstruction with the baseline two-harmonic intrinsic background, $N_\mrm{bkg}=2$.  The additional stellar harmonic absorbs the apparent axion preference seen in the sinusoidal-background fit, yielding a constraint compatible with the triple-dipole result in figure \ref{fig:cornernbkg23dipoles} and with the baseline limit in \eqref{axion_limit}.}
\label{fig:cornernbkg2multipole}
\end{figure}

The result changes once the leading higher harmonic is included in the stellar background.  Figure \ref{fig:cornernbkg2multipole} shows the posterior for $N_\mrm{bkg}=2$.  In this case the apparent preference for nonzero coupling is removed, and the marginalized posterior is concentrated toward small $g_{a\gamma}$, in agreement with the triple-dipole result.  The resulting constraint is compatible with the baseline limit in \eqref{axion_limit}.  The agreement between the two magnetic reconstructions indicates that the main conclusion is not driven by a single parametrization of the PG~1015+014 surface field.  The limiting factor is instead the degeneracy between axion-induced attenuation and intrinsic higher-harmonic stellar variability, which must be marginalized in any phase-resolved photometric search.

\clearpage
\bibliography{ref}
\end{document}